\documentclass[10pt,aps,prb,twocolumn,nisuperscriptaddress, nolongbibliography]{revtex4-2}

\usepackage{amsmath}

\usepackage{lmodern}
\usepackage[T1]{fontenc}
\setcitestyle{super}

\usepackage[utf8]{inputenc}
\usepackage{graphicx}
\usepackage{xcolor}
\usepackage{textgreek}
\usepackage{siunitx}

\usepackage[colorlinks=true,bookmarks=false,citecolor=blue,urlcolor=blue,linkcolor=red]{hyperref}

\begin{document}

\title{On-chip electrically reconfigurable octave-bandwidth optical amplification from visible to near-infrared}

\author{
Guanyu Han$^{1,2}$, Wenjun Deng$^{1,2}$, Yu Wang$^{1,2}$, Ziyao Feng$^{1,2}$ Wei Wang$^{1,2}$, Meng Tian$^{1,2}$, Yu Liu$^{1,2}$, Souvik Biswas$^{3}$, Carlos A. Meriles$^{4,2}$, Andrea Alù$^{1,2}$, and Qiushi Guo${^{1,2,\dagger}}$\\
\vspace{3mm}
\textit{
$^1$Photonics Initiative, Advanced Science Research Center, City University of New York, NY, USA. \\
$^2$Physics Program, Graduate Center, City University of New York, New York, NY, USA.\\
$^3$Department of Electrical and Computer Engineering, University of Michigan, Ann Arbor, MI, USA. \\
$^4$Department of Physics, City College of New York, New York, NY, USA.\\
}
$^\dagger$Email:
\href{mailto:qguo@gc.cuny.edu}{qguo@gc.cuny.edu}
}

\date{\today}

\begin{abstract}
    Achieving broadband on-chip optical amplification spanning the visible and near-infrared (NIR) can enable diverse quantum sensing, metrology, and classical communication applications within a single unified device. However, conventional semiconductor and ion-doped amplifiers suffer from limited gain bandwidths set by fixed energy levels, while optical parametric amplifiers (OPAs) operating continuously from the visible to the NIR have remained elusive due to dispersion-limited bandwidth and the high pump powers required in the visible or ultraviolet (UV). Here, we overcome these limitations by introducing an electrically reconfigurable OPA architecture on lithium niobate integrated photonics. By synergistically combining ultra-high effective  $\chi^{(2)}$ nonlinearity ($\sim$7,000\%/W-cm$^2$), high-order dispersion engineering, and local electro-thermal tuning of quasi-phase matching, our device achieves record gain spectral spanning more than an optical octave, from 770 to 1650 nm. This range covers key transitions of many photonic quantum systems and all telecommunication bands. Moreover, our approach eliminates the need for high-power, wavelength-tunable visible or UV pumps, delivering a peak on-chip gain of 23.67 dB with a single 1060 nm pump at 90 mW average on-chip power. This work opens new avenues for multi-functional, reconfigurable photonics unifying the visible and infrared regimes, with broad implications for quantum sensing and communications. 
\end{abstract}

\maketitle



Despite the significant maturity of photonic integrated circuits (PICs) in telecommunication bands, recent years have witnessed a strong demand for PICs operating over broader optical spectra, especially in the visible and short-NIR regimes\cite{lu2024emerging,chen2026towards,he2023ultra,chauhan2021visible}. This demand is largely driven by the rapid advancements of various atomic and solid-state quantum systems—such as trapped ions, neutral atoms, and solid-state defect centers—which exhibit optical transitions at these shorter wavelengths and play crucial roles in quantum sensing\cite{degen2017quantum,fan2015atom}, time-keeping and navigation\cite{ludlow2015optical,young2020half}, and information processing\cite{levine2018high,omran2019generation}. In these quantum systems, performance is often limited not by intrinsic coherence, but by optical loss, collection inefficiency, and detector noise during readout. Thus, achieving low-noise visible or short-NIR signal amplification can greatly alleviate this bottleneck by boosting weak quantum emitter fluorescence before it encounters lossy or noisy components, enabling higher-fidelity readout, significantly higher quantum sensor sensitivity\cite{lindner2024dual}, and faster measurements. Moreover, a unified, broadband PIC-based amplifier spanning the visible to near-NIR can pre-amplify fragile quantum signals while directly interfacing them with fiber-compatible telecommunication bands, which is essential for low-loss, long-distance quantum networking and distributed quantum sensing and computing\cite{duan2024visible,lu2019efficient,lu2019chip}.

\vspace{0.5mm}

\begin{figure*}[ht]
\centering
\includegraphics[width=0.99\linewidth]{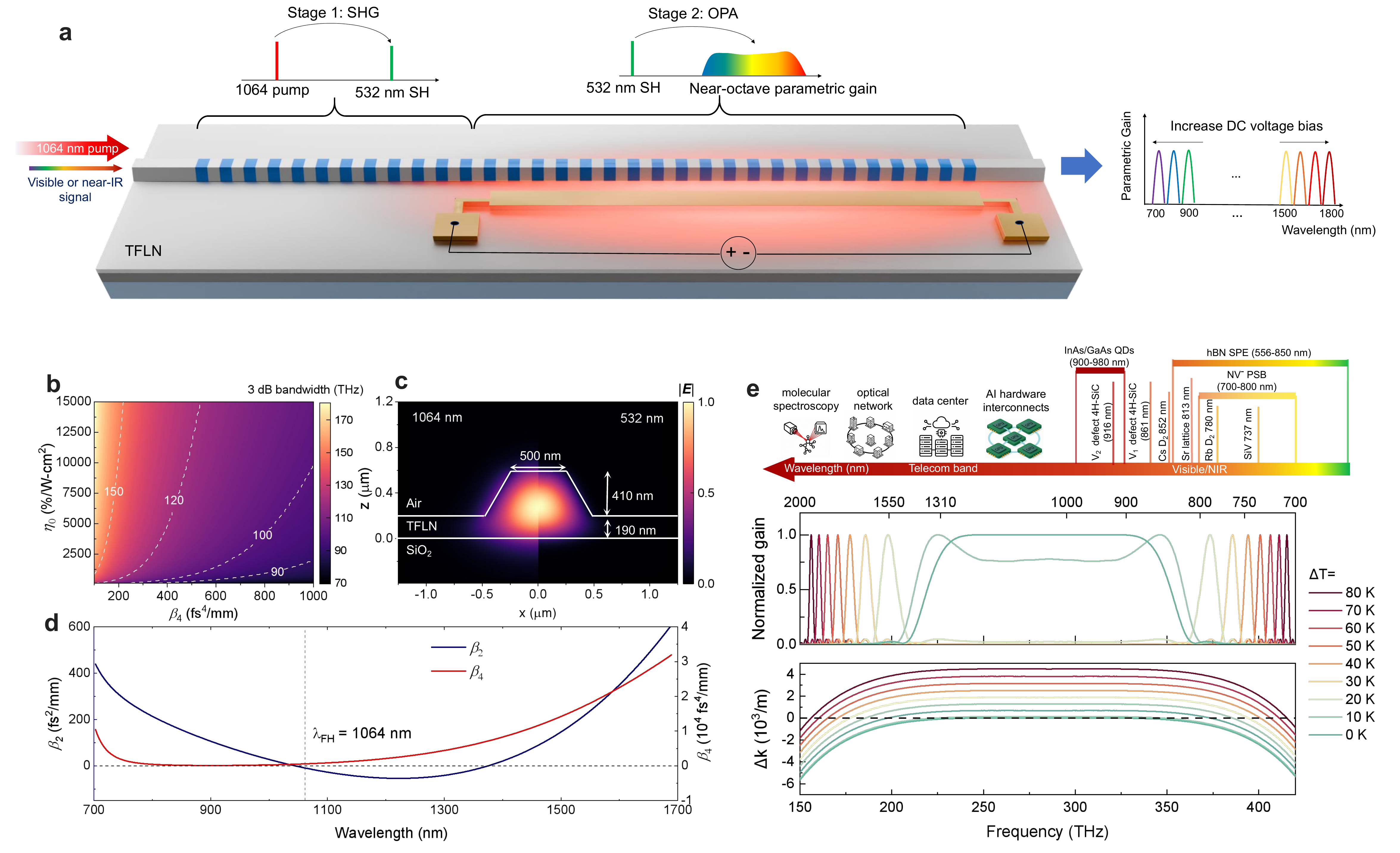}
\vspace{-10pt}
\caption{\textbf{Concept and design of integrated electrically reconfigurable vis-to-NIR OPA.} 
\textbf{a}, Schematic and operating principle of the electrically reconfigurable vis-to-NIR OPA. \textbf{b}, Calculated 3 dB parametric gain bandwidth versus $\eta_0$ and $\beta_4$ when $\beta_2=0$.  \textbf{c}, Optimized waveguide cross-sectional geometry and fundamental TE mode profiles of FH (1064 nm, left) and SH (532 nm, right) inside the waveguide. \textbf{d,} Calculated $\beta_2$ (blue) and $\beta_4$ (red) of the waveguide with optimized cross-sectional geometry, showing a near-zero $\beta_2$ of -3.6 fs$^2$/mm and a low $\beta_4$ of 543 fs$^4$/mm at the FH wavelength. \textbf{e,} Ultra-broadband gain coverage and phase matching characteristics under electro-thermal tuning. At $\Delta T=0$, co-optimization of near-zero $\beta_2$ and $\beta_4$ contributes to a significantly flattened phase mismatch (lower panel) and near-octave gain spectrum near degeneracy (middle panel). Electro-thermal tuning ($\Delta T>0$) further extends the gain spectral overage to 720-1900 nm, supporting various quantum and classical applications across NIR and visible spectrum (upper panel).}
\vspace{-10pt}
\label{Fig1}
\end{figure*}

However, this vision has been thwarted by the bandwidth limitations of conventional optical amplifiers. For example, semiconductor optical amplifiers (SOAs) \cite{van201927, tran2022extending,castro2025integrated,liu2025micro,coldren2012diode} and ion-doped waveguide amplifiers \cite{bao2025erbium,zhang2022chip,yang2024titanium,bradley2011erbium} derive their gain from electronic transitions between fixed material energy levels, thereby restricting their operation to specific, narrow wavelength windows. On the other hand, although optical parametric amplifiers (OPAs) based on cubic ($\chi^{(3)}$) or quadratic ($\chi^{(2)}$) nonlinear parametric process offer spectral flexibility and quantum-limited noise performance\cite{haus2012electromagnetic,kuznetsov2025ultra,dean2025low},  PIC-based OPAs with operational bandwidth from the visible to the NIR have remained elusive to date. Generically, the optical parametric gain is expressed by\cite{boyd2008nonlinear,hansryd2002fiber} $G\propto e^{2gL}$, where $L$ is the interaction length and $g=\sqrt{\Gamma^2-(\Delta k/2)^2}$ is the gain coefficient, dictated by the interplay between the nonlinear drive strength $\Gamma$ and the total phase mismatch $\Delta k$. In $\chi^{(3)}$ Kerr nonlinearity based OPAs, due to the four-wave-mixing (FWM) nature, $\Delta k$ also strongly depends on pump power \cite{hansryd2002fiber}. Although dispersion engineering can broaden the gain spectral window \cite{zhao2025ultra,riemensberger2022photonic,kuznetsov2025ultra,domeneguetti2021parametric,kuyken201150,liu2010mid}, achieving a spectrally flat and broadband gain spectrum is still hindered by such a pump-induced nonlinear phase accumulation\cite{kuyken201150}. Moreover, the intrinsically weak Kerr nonlinearity yields a limited nonlinear drive strength $\Gamma$, which also limits the achievable gain bandwidth. By contrast, $\chi^{(2)}$-based OPAs decouple $\Delta k$ from the pump intensity and benefit from a much stronger $\Gamma$. While these characteristics can in principle enables broader parametric gain bandwidth and higher gain per unit length\cite{jankowski2022quasi,ledezma2022intense,hwang2023mid}, achieving gain in the visible or short-NIR range via parametric down-conversion necessitates short-wavelength (visible/UV) pumps with prohibitively high pump powers. As a result of these challenges, PIC-based OPAs have so far been confined to the telecommunication \cite{riemensberger2022photonic,zhao2025ultra,zhao2025ultra,dean2025low} and mid-infrared \cite{ledezma2022intense,liu2010mid} regimes.  

In this work, we overcome the fundamental bandwidth limitation of conventional optical amplifiers by demonstrating a single-wavelength pumped, electrically reconfigurable $\chi^{(2)}$ OPA on a periodically poled lithium niobate (PPLN) nanophotonic platform. Our device achieves high parametric gain ($>$18 dB) over an unprecedented octave-spanning parametric gain bandwidth from 770 to 1650 nm, effectively bridging the spectral gap between diverse quantum photonic systems and all telecommunication bands. Underlying our innovation is a threefold strategy. First, we utilize an on-chip generated strong visible pump to enable an ultra-high normalized nonlinear conversion efficiency ($\sim$7000\%/W-cm$^2$), which not only provides a strong nonlinear drive for the
OPA but also circumvents the requirement for external
high-power, wavelength-tunable UV/visible pump sources—a longstanding
obstacle for achieving visible-to-NIR OPA. Second, we simultaneously minimize the second- and fourth-order waveguide dispersion, enabling an ultra-broad gain bandwidth at degeneracy. Third, we leveraged the electro-thermal tuning of the  $\chi^{(2)}$ quasi-phase-matching (QPM) condition to further extend the gain spectral coverage, thereby transforming the OPA from a conventionally static device into an electrically wavelength-programmable platform.

\noindent \textbf{Concept of electrically reconfigurable vis-to-NIR OPA.} Figure \ref{Fig1}a depicts the schematic of our electrically reconfigured vis-to-NIR OPA. Built upon a dispersion-engineered PPLN nanophotonic platform, our device uniquely cascades (1) a quasi-phase-matched second-harmonic generation (SHG) section, and (2) a tunable $\chi^{(2)}$ OPA section with electro-thermally tunable phase matching condition. Specifically, when a strong NIR fundamental (FH) light wave centered around 1064 nm (red solid arrow in Fig. \ref{Fig1}a) is injected into the waveguide, it first efficiently generates a co-propagating SH wave at 532 nm and experiences significant depletion. Upon reaching sufficient intensity, the generated 532 nm SH wave then acts as the effective pump for the subsequent tunable OPA process, amplifying the signal wave around the FH wavelength. The advantage of this design is twofold. First, it obviates the need for high power visible pump lasers for pumping the visible-to NIR OPA. Second, the internally generated short-wavelength pump centered around 532 nm yields a significantly enhanced nonlinear drive strength $\Gamma=\sqrt{\eta_0P_{\mathrm{SH}}}$ ($P_{\mathrm{SH}}$ is the power of the generated SH wave) for the OPA process, due to the dramatically enhanced normalized conversion efficiency $\eta_0$ at shorter wavelengths governed by $\eta_0\propto1/\lambda^4$ \cite{jankowski2021dispersion}.

\vspace{1.1mm}

The OPA section is integrated with a metallic microheater for local electro-thermal temperature management. This design translates the OPA into an electrically programmable device, where the phase mismatch $\Delta k$ becomes a function of both signal frequency and the local temperature, i.e. $\Delta k(\omega_\mathrm{s},T)=k({\omega_\mathrm{SH}},T)-k(\omega_\mathrm{i},T)-k(\omega_\mathrm{s},T)-2\pi/\Lambda$,
where $\omega_\mathrm{SH}$, $\omega_\mathrm{s}$, $\omega_\mathrm{i}$ are the angular frequencies of the on-chip generated SH, signal and idler waves, $\Lambda$ is the poling period. The wavevector $k(\omega,T)=n_\mathrm{eff}(\omega,T)\omega/c $  depends on angular frequency $\omega$ and temperature $T=T_0+\Delta T$. Here, $n_\mathrm{eff}$ is the effective refractive index, $c$ is the speed of light in vacuum, $T_0$ is the base temperature and $\Delta T$ is the local temperature change induced by the integrated microheater. At zero voltage bias ($\Delta T=0$), the OPA operates at degenerate regime ($\omega_\mathrm{s}=\omega_\mathrm{i}=\omega_\mathrm{SH}/2$), in which the 3-dB gain bandwidth is given by:
\begin{equation}
\Delta \omega_{\mathrm{3dB}}=\sqrt{\frac{6}{\beta_4}\left (-\beta_2+\sqrt{\beta_2^2+\frac{\beta_4}{3}\sqrt{\frac{4\eta_0P_{\mathrm{SH}}\mathrm{ln2}}{L}}} \right )}
   \label{gainbandwidth}
\end{equation}
where $\beta_2$ and $\beta_4$ represents the second- and fourth-order dispersion, respectively. Clearly, when the second-order dispersion is minimized ($\beta_2\approx0$), $\Delta\omega_\mathrm{3dB}\propto\sqrt[^8]{\eta_0P_\mathrm{SH}/\beta_4^2}$. Figure \ref{Fig1}b shows the calculated 3 dB parametric gain bandwidth by assuming $\beta_2=0$, a $P_\mathrm{SH}$ of 1 W and an interaction length of 5 mm. With a sufficiently high $\eta_0$ together with a low $\beta_4$, the achievable parametric gain around the degeneracy can readily exceed 100 THz.
\vspace{1.0mm}

Motivated by this analysis, we co-optimized $\eta_0$, $\beta_2$ and $\beta_4$ of fundamental TE mode near the FH wavelength around 1064 nm through tailoring the TFLN waveguide cross-sectional geometry. We found that a waveguide top width of 500 nm, an etching depth of 410 nm in a 600 nm-thick TFLN, and a sidewall slant angle of 60° (Fig. \ref{Fig1}c) yield the desired dispersion profile shown in Fig. \ref{Fig1}d, resulting in a near-zero $\beta_2$ of -3.6 fs$^2$/mm and a minimized $\beta_4$ of 543 fs$^4$/mm at 1064 nm. Moreover, as show in Fig. \ref{Fig1}c, with such a waveguide geometry, the fundamental TE modes of the 1064 nm FH and 532 nm SH waves are strongly confined and spatially overlapped, yielding an ultra-high theoretical $\eta_0$ of 11000 \%/W-cm$^2$, calculated by $\eta_0=2Z_0\omega^2d_{\mathrm{eff}}^2/c^2n_{\mathrm{\omega}}^2n_{\mathrm{2\omega}}A_{\mathrm{eff}}$, where $Z_0$ is the impedance of free space, $d_{\mathrm{eff}}=2d_{33}/\pi$ is the effective nonlinear coefficient for a PPLN with an ideal 50\% duty cycle, $n_\omega$ is the effective refractive index of the mode at frequency $\omega$, and $A_{\mathrm{eff}}$ is the effective area\cite{bortz2002noncritical} . Remarkably, this joint optimization enables an ultra-broad near–zero phase-mismatch frequency span, as shown by the lowest green curve in Fig. \ref{Fig1}e lower panel. This leads to an ultra-broad 3 dB parametric gain bandwidth of 132.55 THz (middle green curve in Fig. \ref{Fig1}e middle panel), spanning from 852 to 1367 nm.

\begin{figure*}[ht]
\vspace{-5 pt}
\centering
\includegraphics[width=0.9\linewidth]{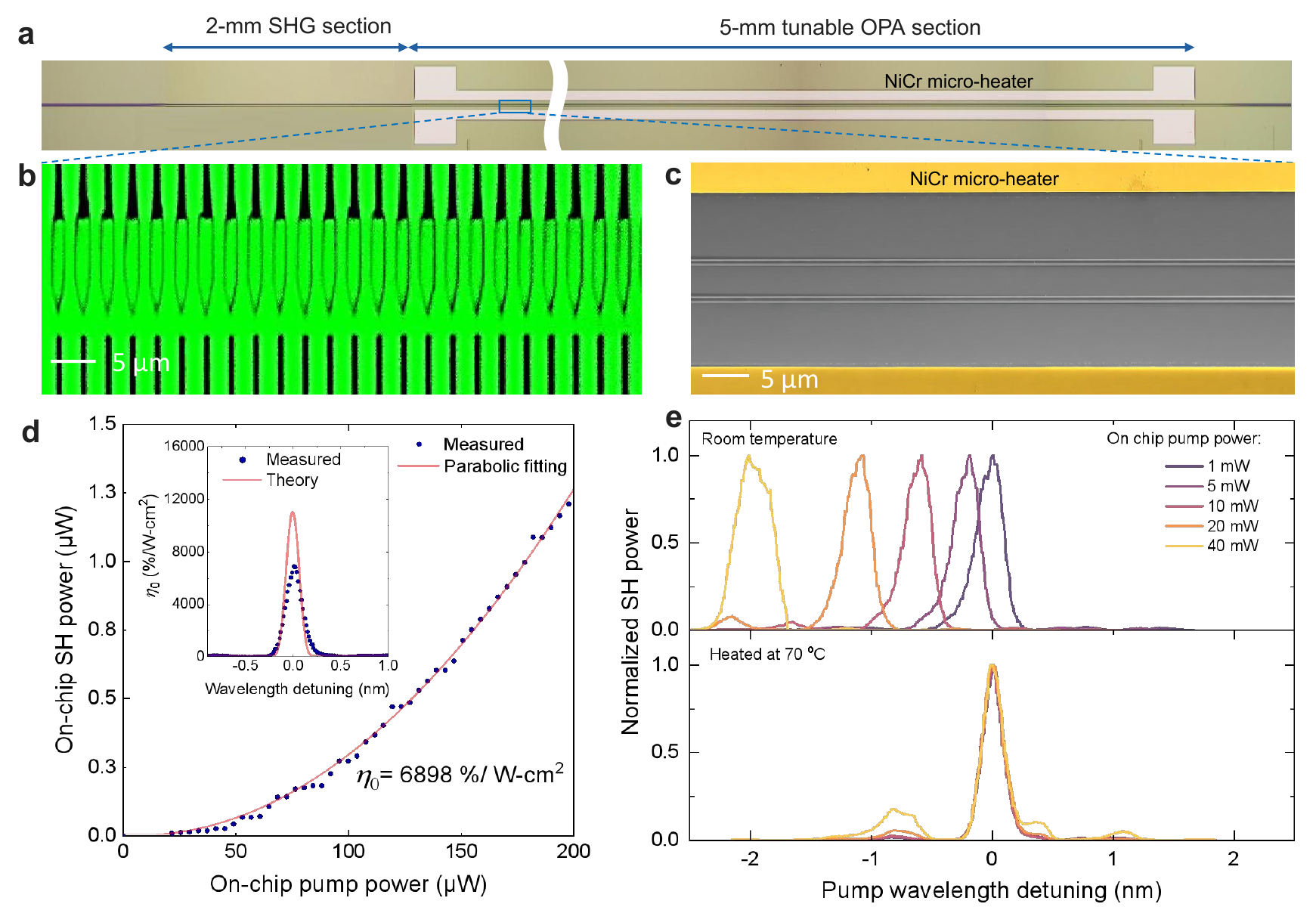}
\vspace{-5 pt}
\caption{\textbf{Integrated electrically reconfigurable vis-to-NIR OPA and its basic $\chi^{(2)}$ nonlinear characteristics. } \textbf{a}, Optical microscope image of the OPA, which consists of a 2-mm-long SHG section followed by a 5-mm-long tunable OPA section. \textbf{b}, Two-photon microscopy of the periodic poling before waveguide fabrication, revealing a poling period of 2.55 $\mu$m with a duty cycle of $\sim$50\%. \textbf{c}, 45$^{\circ}$ titled false color SEM image of the nanophotonic TFLN waveguides and neighboring NiCr microheaters (yellow). Two waveguides were fabricated within one poling region. \textbf{d}, Measured on-chip SH power (blue symbols) as a function of FH power and its parabolic fit (red solid line), showing an ultra-high $\eta_0$ of 6898 \%/W-cm$^2$. Inset: Measured (blue) and theoretical (red) SHG power as a function of the input FH wavelength. \textbf{e}, Upper panel: At room temperature, increased pump power induces a strong photorefractive blue-shift of the phase-matching wavelength. Lower panel: Elevating the chip temperature ($T_0$) to 70 °C yields a power-independent, stable phase-matching wavelength.}
\vspace{-10 pt}
\label{Fig2}
\end{figure*}
 
Applying a DC voltage bias to the metallic micro-heater induces a local temperature rise ($\Delta T>0$) of the OPA section. While the phase matching condition for the SHG section remains unchanged, that of the OPA section gets perturbed due to thermo-optic effect, leading to a shift in the frequencies of peak parametric gain in accordance with energy conservation. As shown in the bottom panel of Fig. \ref{Fig1}e, the phase-mismatch spectrum translates vertically as $\Delta T$ increased from 0 to 80 K. This upward shift causes the phase-matching points ($\Delta k=0$) to shift away from the degeneracy frequency, driving the OPA into non-degenerate regime, where the gain spectrum splits into two voltage-programmable sidebands (Fig. \ref{Fig1}e middle panel). The sensitivity of phase matching frequency $\omega_\mathrm{s}$ to temperature variations can be described by the relation:
\begin{equation}
   \frac{d\omega_\mathrm{s}}{dT}\propto-\frac{1}{(\beta_\mathrm{1s}-\beta_{\mathrm{1i}})-(\beta_{\mathrm{2s}}+\beta_{\mathrm{2i}})\Delta\omega}
\end{equation}
Here, $\Delta\omega=|\omega_\mathrm{s}-\omega_{\mathrm{FH}}|$ is the signal frequency detuning from the FH, $\beta_{\mathrm{1s}}-\beta_{\mathrm{1i}}$  represents the group velocity mismatch (GVM) between signal and idler wave and $\beta_{2\mathrm{s,i}}$ is the second-order dispersion for signal and idler waves, respectively. Near the degeneracy, where $\beta_{\mathrm{1s}}-\beta_{\mathrm{1i}}\approx0$, the near-zero $\beta_2$ significantly enlarges the tuning slope. As the detuning increases, the growing GVM gradually reduces tuning sensitivity. As shown in Fig. \ref{Fig1}e middle and bottom panels, this mechanism facilitates an efficient, beyond-octave gain spectral tuning from 720 nm 1900 nm with a modest $\Delta T$ of 80 K. By dynamically aligning the high-gain windows with specific optical transitions, our OPA can be tailored to amplify diverse targets ranging from visible quantum emitters, such as SiV centers (737 nm) and Rb D$_2$ lines (780 nm), to telecommunication-band signals marked by Fig. \ref{Fig1}e upper panel.

\begin{figure*}[ht]
\centering
\includegraphics[width=0.97\linewidth]{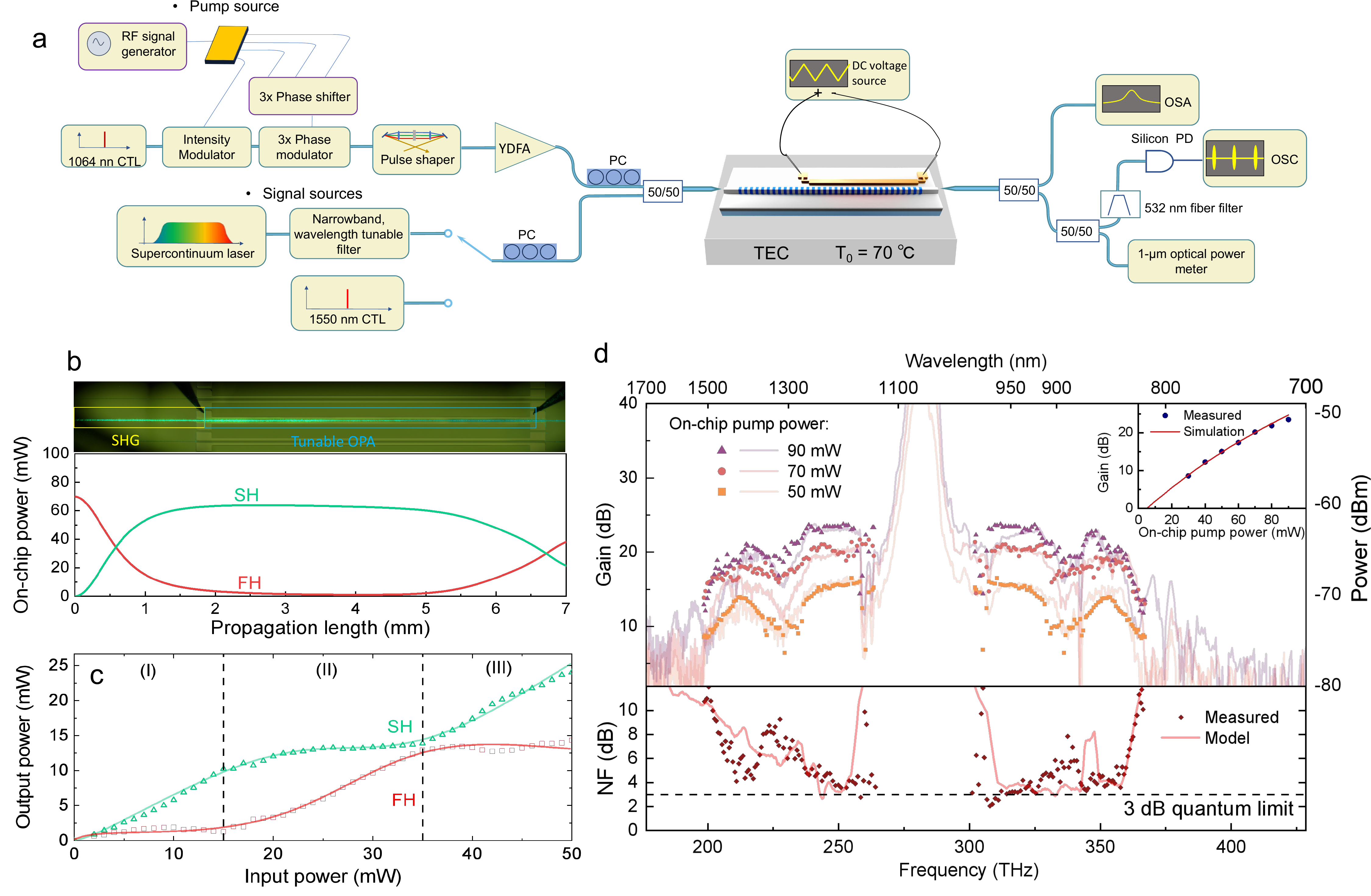}
\vspace{-5 pt}
\caption{\textbf{Characterization of OPA at degeneracy ($V_\mathrm{DC}=0$).} \textbf{a}, Experimental setup for OPA measurements. The OPA chip is mounted on a TEC stablized at $T_0=70^\circ\mathrm{C}$. A programmble DC bias is applied for thermo-optic tuning. CTL: continuous tunable laser; YDFA: ytterbium-doped fiber amplifier; TEC: thermoelectric cooler; OSA: optical spectrum analyzer; OSC: oscilloscope; PC: polarization controller; PD: photodetector. \textbf{b}, Upper panel: microscope image of the OPA showing the spatial evolution of the on-chip generated (green light) SHG power. Lower panel: Simulated on-chip FH (red) and SH (green) power evolution for an on-chip average pump power of 70 mW. \textbf{c}, Measured on-chip SH and FH output powers (symbols) versus on-chip input FH average power, compared with simulation (solid lines). \textbf{d}, Upper panel: measured parametric gain (symbols) for on-chip pump powers of 50, 70, and 90 mW (with $V_\mathrm{DC}=0$), overlaid with the corresponding spontaneous parametric fluorescence (solid lines). Inset: measured peak gain (blue symbols) at 920 nm as a function of on-chip pump power, showing good agreement between experiment and simulation (red solid line). Lower panel: measured (symbol) and calculated (solid line) wavelength-dependent noise figure (NF) (symbols) with 90 mW on-chip FH average power.}
\label{Fig3}
\vspace{-10 pt}
\end{figure*}

\vspace{1.3mm}
\noindent\textbf{$\chi^{(2)}$ nonlinear property characterization and device stabilization.} Figure \ref{Fig2}a shows the optical microscope image of the integrated OPA (see Methods for device fabrication details). The device comprises a 7-mm-long periodically PPLN nanophotonic waveguide with a uniform poling period and an optimized waveguide cross-section shown in Fig.\ref{Fig1}c. The first 2 mm of the PPLN waveguide serves as an SHG section, followed by a 5-mm-long tunable OPA section. The quality of periodic poling structure before waveguide fabrication can be seen in Fig. \ref{Fig2}b through two-photon microscopy, revealing uniform ferroelectric domains with a nearly ideal 50\% duty cycle and a poling period of 2.55 $\mu$m. In addition, a 5-mm-long NiCr micro-heater was fabricated alongside the OPA section at a distance of 10 $\mu$m from the waveguide, as shown in the false color scanning electron microscope (SEM) image (Fig. \ref{Fig2}c).

\vspace{1.3mm}
\begin{figure*}[ht]
\vspace{-5 pt}
\centering
\includegraphics[width=1\linewidth]{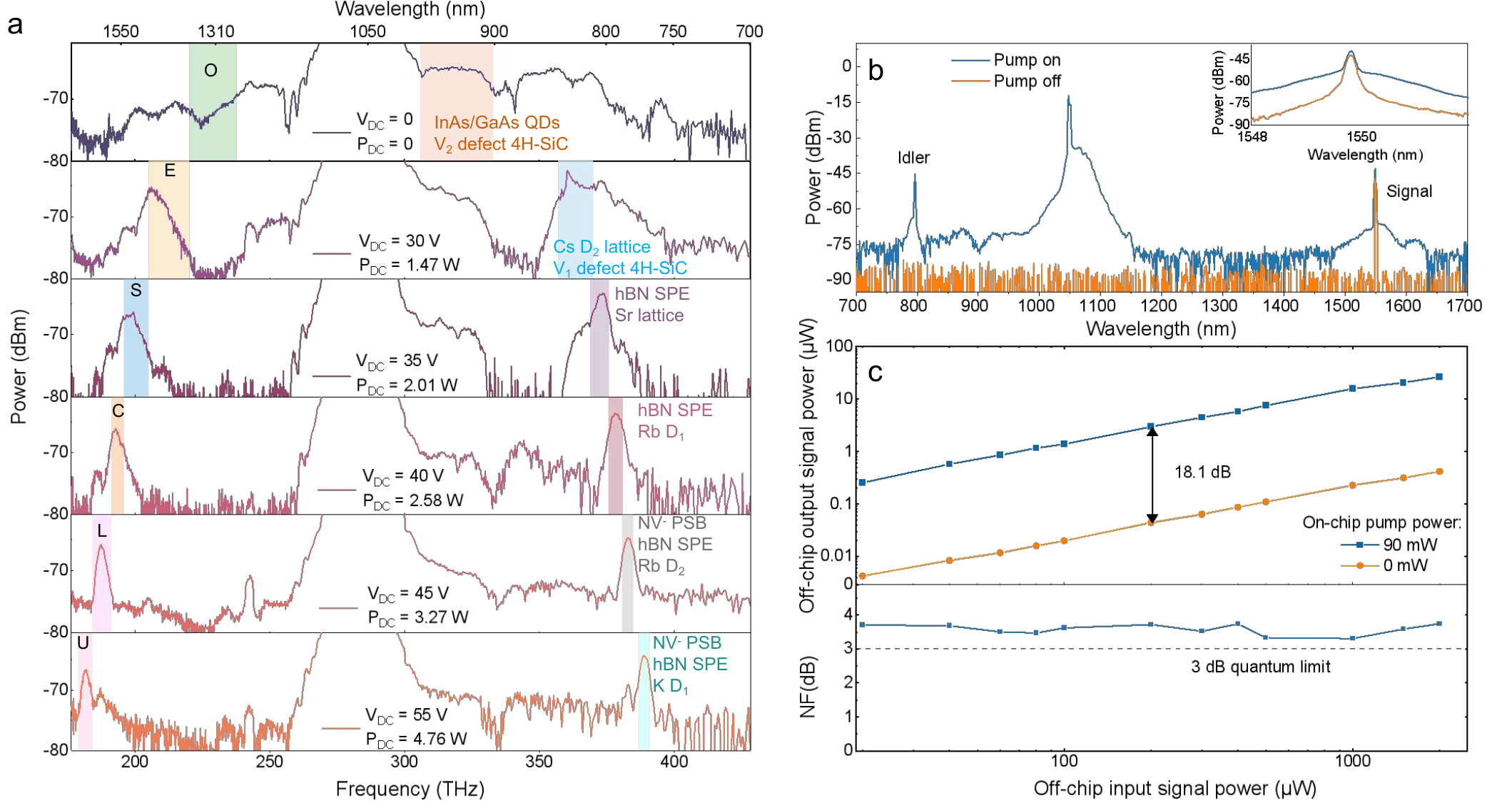}
\vspace{-5 pt}
\caption{\textbf{Electrically reconfigurable visible-to-NIR OPA.} \textbf{a}, Electrically reconfigurable parametric gain across visible and NIR bands at different V$_\mathrm{DC}$. Upon applying V$_\mathrm{DC}$, the OPA switches to the nondegenerate regime and the gain spectral coverage is further extended to more than one octave, from 770 nm to 1650 nm. Shaded areas highlight telecommunication bands and important spectral regions for photonic quantum technologies. QD: quantum dot; hBN: hexagonal boron nitride; SPE: single-photon emitter; PSB: photon sideband. \textbf{b}, output optical spectra measured without (orange) and with (blue) 90 mW on-chip average power at 1060 nm when an 2 mW (off-chip power) 1550 nm coherent signal was injected.
Inset: Zoomed-in optical spectra. The span of the OSA is 4 nm with a 0.05 nm resolution. \textbf{c}, Upper panel: on-chip signal output power at wavelength 1550 nm as a function of signal powers at the input of the OPA, showing an unsaturated on/off gain of 18.1 dB at a 90 mW on-chip pump power. Lower panel: measured NF as a function of on-chip signal power. } 
\label{Fig4}
\vspace{-10pt}
\end{figure*}

First, we characterize the $\chi^{(2)}$ nonlinear property of the device by measuring the SHG spectral response and its efficiency with low-power input FH around 1064 nm. As shown in the inset of Fig. \ref{Fig2}d, the measured SHG transfer function exhibits a near sinc-shaped profile. We then measured the output SH power as a function of input FH power at the phase matching wavelength. Figure \ref{Fig2}d shows that the measured on-chip SH power scales quadratically with the fundamental pump power in the undepleted regime, from which we extracted an ultra-high $\eta_0$ of 6,898 \%/W-cm$^2$. Compared to the theoretical predicted $\eta_0$ of 11,000 \%/W-cm$^2$, the lower $\eta_0$ and slightly distorted SHG spectral features (broader sinc profile and elevated side-lobes) are mainly attributed to inhomogeneities in the thin-film thickness throughout the 7-mm-long PPLN waveguide. Under higher input FH power around 1064 nm, the generated SH around 532 nm induces significant photorefractive effect in the waveguide\cite{kostritskii2009photorefractive}, leading to a pronounced blue-shift of the phase-matching wavelength and broadening of sinc-shaped SHG response at room temperature, as shown in the upper panel of  Fig. \ref{Fig2}e. Such a power-dependent operating point shift is detrimental to the reliable operation of the OPA. To mitigate this effect, we elevated the global OPA chip temperature ($T_0$) to 70°C using a thermoelectric cooler (TEC), which is known to greatly suppress the photorefractive instability by increasing the photo-conductivity to counteract the space-charge field\cite{rams2000optical}. This approach effectively yields an almost power-independent phase-matching wavelength, as depicted in the lower panel of Fig. \ref{Fig2}e. 
\vspace{1.0mm}

\noindent\textbf{OPA Measurements.} The experimental setup used for measuring the electrically reconfigurable OPA is shown in Fig. \ref{Fig3}a. The pump source consists of a 7 GHz, $\sim$3 ps pulse train with center wavelength tunable around 1064 nm, generated from a 1050 nm continuous-wave tunable laser (1050 CTL) followed by an electro-optic frequency comb system (see methods for details). To quantify the broadband parametric gain, we employed spectrally filtered and tunable supercontinuum source or a wavelength-tunable CW laser around 1550 nm as the signal sources. The device is mounted on a TEC and stabilized at a temperature of 70 °C. A DC voltage ($V_{\mathrm{DC}}$) supplied by a source meter was applied to the device through electrical probes, enabling electro-thermal tuning of the OPA gain spectrum.
\vspace{1.0mm}

We first perform the high-power input-output measurement at $V_{\mathrm{DC}}=0$ in the absence of input signal. Figure \ref{Fig3}b upper panel shows the optical microscope image of the OPA under a high on-chip average input FH power of 70 mW, which translates to an on-chip peak power of 2.94 W assuming a sech$^2$ pulse shape. At the phase matching wavelength ($\lambda_\mathrm{FH}=$1061 nm), the on-chip generated SH (green light) power grows rapidly within the first 2-mm SHG section and is subsequently attenuated in the OPA section. This behavior is corroborated by numerical simulated SH and and FH power along the waveguide based on a single-envelope model (Fig. \ref{Fig3}b, lower panel). When the generated SH power in the SHG section becomes sufficiently strong, it significantly depletes the FH. Then, the intense SH field, together with the input noise, spontaneously initiate the subsequent OPA process, in which the generated SH acts as the pump and transfers energy to the FH and spectral components around it. 

Figure \ref{Fig3}c plots the measured on chip SH (green symbols) and residual FH (red symbols) output powers as functions of the on-chip input FH power. The measurement results agree well with the simulations (solid lines) and clearly reveal three distinct regimes. When the input FH power is weak ($<$15 mW, regime-I), the device operates in the SHG regime, where the SH output increases with a reduced slope and the FH is strongly depleted. As the input FH power exceeds 15 mW (regime-II), the internally generated SH becomes sufficiently intense to initiate the OPA process, leading to significant saturation of SH power growth and a simultaneous resurgence of the FH. In the high-power regime ($>$ 35 mW, regime-III), a resurgence in the SH power growth is observed, indicating repeated energy exchange between FH and SH. In this high-power regime, vacuum fluctuations and noise components carried by the input FH source across the 800–1500 nm spectral range are strongly amplified to macroscopic levels, giving rise to the ultra-broadband parametric fluorescence shown in Fig. \ref{Fig3}d (solid lines).

To characterize the broadband parametric gain at $V_{\mathrm{DC}}$= 0, we injected a probe signal from a supercontinuum source and continuously tuned its spectral passband from 820 to 1500 nm (See methods for details), with the off-chip average power fixed at 1 $\mu$W. The upper panel of Fig. \ref{Fig3}d shows the measured on/off gain (symbols) as a function of frequency under various on-chip FH average powers. The measured gains at degeneracy appear to be ultra-broadband (858 to 1480 nm), with spectral features agreed well with the independently measured spontaneous parametric fluorescence spectra (solid lines). As shown in the inset of Fig. \ref{Fig3}d, increasing the on-chip FH pump power results in a monotonic increase in the on/off gain, reaching a peak value of 23.67 dB at 920 nm under 90 mW of on-chip FH average power. Moreover, the FH pump-power dependence of the peak gain at 920 nm is well reproduced by our single-envelope model. In addition, we evaluated the wavelength-dependent noise figure (NF) under an on-chip FH average power of 90 mW (Fig. \ref{Fig3}d, lower panel) using the relation $\mathrm{NF}=2\rho_{\mathrm{ASE}}/{G_{\mathrm{net}}h\nu}+1/{G_{\mathrm{net}}}$\cite{baney2000theory}, where $G_{\mathrm{net}}$ is the net gain and $\rho_{\mathrm{ASE}}$ is the parametric fluorescence level with the assumption of a bandwidth of 0.01 nm. As shown in Fig. \ref{Fig3}d lower panel, the noise figure (NF) approaches the 3-dB quantum limit for phase-insensitive amplification over three spectral windows, spanning 840–876 nm, 920–1239 nm and 1381–1421 nm. Outside these bands, the NF increases primarily because the parametric gain drops so that fixed optical losses and measurement noise floor contribute more strongly to the extracted NF. The solid curve is calculated from the measured parametric fluorescence spectrum in the top panel, which shows a consistent comparison between the measured gain and inferred noise performance.
\vspace{1.1mm}

Furthermore, we demonstrate the electrical reconfigurability of the OPA gain spectra by applying various $V_{\mathrm{DC}}$. In our device, the local temperature rise ($\Delta T$) of the OPA waveguide scales linearly with the electrical power by 19.4 K/W, which was calculated through the 2-D heat transfer simulation. Figure \ref{Fig4}a shows the measured parametric fluorescence spectra at several $V_{\mathrm{DC}}$ and corresponding electrical powers. Increasing the $V_{\mathrm{DC}}$ does not perturb the SHG section, but strongly reconfigures the phase-matching condition of the tunable OPA section, shifting the OPA gain window across a wide spectral range from 770 nm to 1650 nm. In particular, it covers all major telecommunication bands, including the O-band for emerging co-packaged optical interconnects \cite{ahmed2025universal,hua2025integrated} and low-dispersion fiber links, as well as the C-band used for low-loss long-haul telecommunication links and numerous critical wavelengths associated with solid-state single-photon emitters and atomic transitions.
\vspace{1.1mm}

Finally, we quantified the gain after electro-thermal tuning. Here, the $V_{\mathrm{DC}}$ was set to 40 V to shift the gain peak to $\sim$1550 nm. As shown in Fig. 4b, with a 2 mW off-chip input signal at 1550 nm, activating the 90 mW, 1061 nm FH pump produces a 3.43 dB signal enhancement measured on the optical spectrum analyzer (OSA), accompanied by a strong idler peak at 796 nm. This time-averaged value, however, significantly underestimates the true parametric gain due to the pulsed pump. The instantaneous peak power of the amplified CW signal is extracted as  $P_\mathrm{peak,on}=\int P_\mathrm{out}(\lambda)d\lambda/(f_\mathrm{rep}\tau_{\mathrm{p}})$, where $P_\mathrm{out}(\lambda)$ is the average output power spectral density of the signal, and $f_\mathrm{rep}$ and $\tau_{\mathrm{p}}$ are the repetition rate and pulse width of the input FH pulse. The resulting peak gain is 18.1 dB, calculated by $G(\mathrm{dB})=10\log_\mathrm{10}(P_\mathrm{peak,on}/P_\mathrm{off})$. We further measured the output power as a function of input signal power by varying the off-chip signal power from  20 $\mu$W to 2 mW. As shown in the upper panel of Fig. \ref{Fig4}c, the output power scales almost linearly with the input over the measured range, indicating negligible gain saturation. In addition, the on-chip NF, calculated following the procedure described in Methods, is plotted in the lower panel and remains close to the 3 dB phase-insensitive quantum limit. 
\vspace{1.2mm}

\begin{figure}[ht]

\centering
\includegraphics[width=1\linewidth]{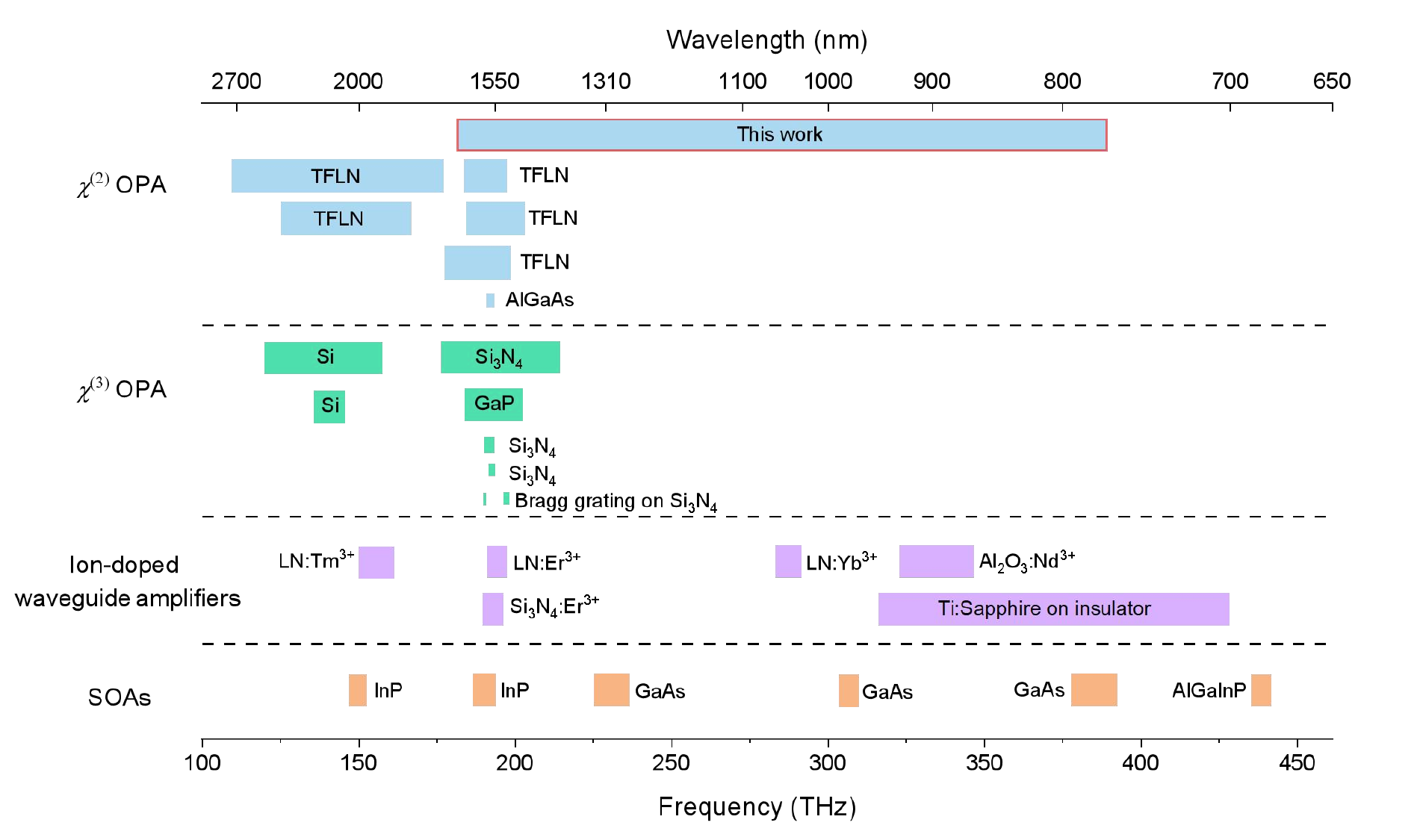}
\caption{\textbf{Comparison of operational bandwidth of various optical amplifier technologies.}
}\label{Fig5}
\vspace{-10 pt}
\end{figure}

\noindent\textbf{Discussion and outlook.} Figure \ref{Fig5} highlights that the demonstrated device (red framed blue box) occupies a parameter space that is not accessible to conventional PIC-based optical amplifiers. While SOAs based on III–V semiconductors can be engineered for particular spectral windows from the visible to the near-IR\cite{van201927,franken2021hybrid,tran2022extending,castro2025integrated,liu2025micro,volet2017semiconductor}, each individual device is fundamentally constrained by its material bandgap and carrier distribution, limiting its usable bandwidth when multiple, widely separated wavelengths must be addressed simultaneously. Similarly, solid-state ion-doped waveguide amplifiers exhibit discrete gain bands dictated by the fixed energy levels of the dopant ions, for example Er$^{3+}$ in the C/L bands of 1550 nm \cite{liu2022photonic, zhou2021chip} and Ti:sapphire in the 700-980 nm range \cite{yang2024titanium,wang2023photonic}.  OPAs exploit parametric nonlinear interactions mediated by virtual energy levels to transfer energy from a pump to a signal, and are therefore not constrained by material-defined electronic transitions.  However, to date, no PIC-based OPAs have operational gain windows below 1 $\mu$m, nor achieved spectral coverage exceeding one octave, which are capabilities demonstrated in this work.

Our reconfigurable OPA opens new pathways toward wavelength-agile light sources for high-resolution visible atomic and molecular spectroscopy\cite{bohn2017cold}, as well as visible-to-telecom quantum transduction\cite{rakher2010quantum,guo2016chip,lu2019chip,lu2019efficient,wang2022synthetic,duan2024visible}. It also enables emerging opportunities for parametric-gain-enhanced quantum sensing and information processing in atomic systems and solid-state quantum defects, where optical signals are often weak and noisy. For example, by tuning the OPA gain window to defect emission wavelengths, our device can act as a pre-amplifier for weak fluorescence prior to detection, fiber coupling or on-chip routing, improving readout fidelity and reducing integration times in quantum processors and sensors. Moreover, when integrated with an on-chip optical resonator, our OPA could enable transition-edge–type quantum sensors with dramatically enhanced sensitivity\cite{webb2021laser,jeske2016laser,lindner2024dual}, converting weak electromagnetic perturbations of quantum systems into large output optical intensity variations while simultaneously generating telecom-band idler photons for long-distance sensor interconnects\cite{lu2019efficient}.
\vspace{1.1mm}

Looking forward, several avenues exist to further enhance the performance and versatility of our reconfigurable OPA architecture. The gain spectral ripples shown in Fig. \ref{Fig3}d, attributed to LN thin-film thickness variations, can be mitigated by implementing the adaptive poling technique\cite{chen2024adapted,li2025integrated}. Crucially, adaptive poling also enables the use of long meandering PPLN waveguides for further enlarging the nonlinear interaction length. Extending the interaction length, combined with improvements in fiber-to-chip coupling \cite{hu2021high,guo2024polarization} and the use of intra-cavity second-harmonic resonances\cite{dean2025low}, creates a promising pathway of achieving high-gain with a modest CW pump power.  In addition, the spectral reach of our OPA in the visible can be further extended by shifting the pump wavelength to the 900–980 nm range, which further reduces $\beta_4$ and thereby maximizes both the gain bandwidth at degeneracy and the overall tuning range. Furthermore, the spectral tuning range and efficiency can be significantly improved by integrating micro-heaters in closer proximity to the waveguides\cite{li2024advancing} and by leveraging suspended TFLN structures\cite{liu2022thermally}. We anticipate that these advances will establish our reconfigurable OPAs as a key building block for fully programmable, spectrum-spanning photonic systems, supporting next-generation quantum and classical photonic applications.

\section*{Methods}
\noindent \textbf{Device fabrication.} The OPAs were fabricated on a 600-nm-thick, X-cut, magnesium-oxide (MgO)-doped thin-film lithium niobate (TFLN) layer bonded to a SiO$_2$ (4.7 $\mu$m) / silicon (500 $\mu$m) substrate. Fabrication began with the definition of periodic-poling electrodes. Electrode patterns were written using 50-keV electron-beam lithography (Elionix ELS-G50) with polymethyl methacrylate (PMMA) as the resist, followed by electron-beam evaporation of 15 nm Cr and 55 nm Au and subsequent lift-off in acetone. Periodic poling was performed in oil by applying a 500 V peak voltage pulse (100 $\mu$s period, 50\% duty cycle) to the electrodes. After poling, the Cr/Au electrodes were removed using standard metal etchants. The waveguides were then defined using 100-keV electron-beam lithography (Elionix ELS-G100) with hydrogen silsesquioxane (HSQ) as the resist. An Ar$^+$-based inductively coupled plasma (ICP) etch was used to etch 410 nm into the TFLN film, producing waveguides with $\sim$ 60° sidewall angles. The residual HSQ mask was subsequently removed using a buffered oxide etchant (BOE). A 200-nm-thick NiCr microheater was fabricated using the same lithography and lift-off process as the periodic-poling electrodes. Finally, the input and output facets of the waveguides were mechanically polished to improve fiber-to-chip coupling efficiency. 

\vspace{1.3mm}

\noindent \textbf{Optical measurements.} The SHG efficiency and spectral response shown in Fig. \ref{Fig2}d and e were characterized by pumping the device with CW light from a continuous tunable laser (Topica CTL 1050). The pump light was coupled into the waveguide via a single-mode lensed fiber. The output light was collected by another single-mode lensed fiber. To measure the SH power exclusively, the remaining FH was filtered out by a fiber filter which has a passband below 700 nm. The filtered SH light was directed to a silicon photodetector, and the resulting photocurrent was measured using an oscilloscope to determine the SH output power. The input/output coupling loss were estimated to be 10.5 dB and 11.7 dB.  

We characterized the gain performance of the OPA using a pulsed pump, with a 1061.62 nm center wavelength, 7 GHz repetition rate and 3 ps pulse width. This pump was generated by an electro-optic (E-O) comb configuration (Fig. \ref{Fig3}a), where the CW light was modulated by one intensity and three phase modulators, which were driven by an amplified RF signal. A programmable pulse shaper (Coherent Waveshaper 1000A/SP) was implemented to compress the pulses to 3 ps by setting appropriate dispersion. A high-power (33 dBm) YDFA (Civil Laser) was used to amplify the pump to the required power level. For broadband gain characterization, we used two different seed lasers: a supercontinuum source (NKT Photonics, SuperK FIANIUM) in conjunction with an acousto-optic filter and a CW laser centered at 1550 nm. To monitor the input powers, 1\% of both the pump and the selected seed laser were tapped via 99:1 splitters and sent to a power meter (Thorlabs PM20). The remaining pump and seed were then combined using a 50:50 coupler and coupled into the waveguide with a lensed fiber. During measurement, the chip was placed on a TEC and the temperature was stabilized to 70 °C with a temperature controller (Vescent SLICE-QTC ) and a thermistor. To map the broadband gain, we swept the passband of the filter of the supercontinuum laser from 820 nm to 1500 nm in 5 nm steps, maintaining a fixed off-chip signal power of 1 $\mu$W. The output spectra for both pump-on and pump-off cases were recorded using an optical spectrum analyzer (ANDO AQ6315A) with a 0.05 nm resolution.
\vspace{1.3mm}

\noindent\textbf{Peak gain calculation.} 
The large disparity in repetition rates and pulse widths between the pump and signal from the supercontinuum laser led to a significant underestimated gain when measuring the signal enhancement using the OSA. Assuming a gaussian profile, the on-chip generated SH pump pulse width is $\tau_{\mathrm{SH}}=3  \mathrm{ps}$, while the signal pulses are 350 ps at 78 MHz. Since the signal pulse width is significantly longer than the SH pump period ($T_{\mathrm{SH}}\approx143$ ps) each signal pulse temporally overlaps with approximately $N=\tau_{s}/T_{\mathrm{SH}}\approx2.45$ SH pump pulses. Since the amplification only occurs during the transient windows defined by the SH pump pulse width, the measured average gain $G_{\mathrm{avg}}$, which was obtained by comparing the OSA-reported value, underestimates the true peak gain $G_{\mathrm{peak}}$ according to the following temporal duty cycle correction:  $G_{\mathrm{peak}}=1+(G_{\mathrm{avg}}-1)/F$, where the temporal overlap $\eta$ is defined by the SH pump duty cycle: $F=f_{\mathrm{rep}}\cdot\tau_{\mathrm{SH}}\approx0.021$. Under high gain regime ($G_{\mathrm{peak}}\gg1$), the actual peak gain is is approximately 16.8 dB higher than the observed average gain.

\vspace{1.3mm}

\noindent \textbf{Numerical simulations.} We used a commercial software (Ansys Lumerical Inc.) to solve for the waveguide modes and the dispersion properties. In the simulation, the anisotropic refractive index was modeled by the Sellmeier equations\cite{zelmon1997infrared} . For the nonlinear optical simulation, we numerically solved an analytical nonlinear envelope equation (NEE) in the frequency domain using a split-step Fourier technique to simulate the pulse propagation and nonlinear dynamics in the waveguide \cite{phillips2011supercontinuum}. The nonlinear step was solved with a fourth-order RungeKutta method. We used the COMSOL for calculation of thermal performance of the NiCr microheater and the resulting localized heating of the TFLN waveguide. 


\section*{Data Availability}
The data that support the plots within this paper and other findings of this study
are available from the corresponding author upon reasonable request.
\section*{Code Availability}
The computer code used to perform the nonlinear optical simulations in this paper is available from the corresponding author upon reasonable request.
\section*{Acknowledgments}
The authors acknowledge support from NSF Grant No. 2338798 (NSF CAREER Award), the start-up grants from the CUNY Advanced Science Research Center and the CUNY Graduate Center, and the PSC CUNY Research Award. The authors thank Prof. Gabriele Grosso and Dr. John Woods for providing access to laboratory facilities, and Dr. Dmitriy Korobkin and Dr. Viktoriia Rutckaia for technical assistance.

\section*{Authors Contributions}
Q.G. conceived the idea and supervised the project. G.H. and W.D. designed the devices. G.H. fabricated the devices and performed the measurements with assistance from W.W., W.D., and Z.F. G.H. performed the theoretical modeling and simulations with assistance from  W.D. and Y.W. G.H. and Q.G. wrote the manuscript with input from all authors.
\section*{Competing Interests}
The authors declare no competing interests.

\bibliographystyle{apsrev4-2}

\bibliography{references}

\end{document}